\documentclass[11pt]{article}
\usepackage{graphicx} 
\usepackage[margin=1.5in]{geometry} 

\usepackage{amsmath}
\usepackage{authblk}
\usepackage[backend=biber, sorting=none]{biblatex}
\title{\textbf{Fast thermo-optic switching on silicon nitride platform through parity-time symmetry breaking}}

\author[]{\textit{Ravi Pradip}}
\author[]{\textit{Venkata Sai Akhil Varri}}
\author[]{\textit{Liam McRae}}
\author[]{\textit{Frank Brückerhoff-Plückelmann}}
\author[]{\textit{Daniel Wendland}}
\author[]{\textit{Anna P. Ovvyan}}
\author[]{\textit{Shabnam Taheriniya}}
\author[]{\textit{Wolfram Pernice}}
\author[]{\textit{Simone Ferrari}}

\affil[]{\small Kirchhoff-Institute for Physics, University of Heidelberg, Germany}

\date{} 

\begin{document}
\maketitle

\begin{abstract} 
\textbf{This work demonstrates a fast thermo-optic switching mechanism on silicon nitride on insulator platform leveraging parity-time symmetry breaking. The cladding-free design enables low-loss optical propagation in a partially metal-covered waveguide, with the same metal layer serving as an integrated heater for rapid phase tuning. The fabricated device exhibits an 8.5 $\mu$s rise time for a $\pi$ phase shift, despite the weak thermo-optic coefficient in silicon nitride. Additionally, the impact of thermal cross-talk is investigated and an insertion loss as low as 0.39 dB for a 100-$\mu$m-long heater-waveguide section is demonstrated.}
\end{abstract}

\section{Introduction}

Silicon Nitride (\(\text{Si}_3\text{N}_4\)) on insulator emerges as a highly versatile platform for advanced integrated photonic circuits within the field of CMOS technology \cite{sin1, sin2}. Its exceptional properties include a broad transparency range extending from the visible to the mid-infrared spectrum (470–6700 nm) \cite{SiN_vis}, coupled with remarkably low propagation losses down to 0.1 dB/m \cite{ultralowloss}. These attributes make it particularly advantageous for various applications, encompassing telecommunications\cite{SiNTele}, life sciences \cite{SiNLifescience}, LiDAR \cite{SiNLidar}, and photonic computing \cite{Xiang:22}. As the demand for advanced functionalities in nanophotonic devices continues to rise, the ability to dynamically reconfigure these systems is becoming increasingly crucial. 

\( \text{Si}_3\text{N}_4 \) is a material that lacks intrinsic free carriers and does not exhibit an inherent electro-optic effect. As a result, modulation strategies on this platform are typically reliant on methods such as charge carrier injection \cite{doping} or multi-layer heterogeneous integration \cite{Churaev2023}. Although fast-responding devices with bandwidths exceeding several tens of gigahertz have been achieved through these methods, they are often associated with high losses or add complex steps to the fabrication procedure, making their implementation challenging. On the other hand, for applications with modest bandwidth requirements, in the range of several tens of kilohertz, the thermo-optic (TO) effect offers a highly scalable and cost-effective solution. Tunable devices based on the TO effect, which leverages the temperature-dependent refractive index of materials, are particularly advantageous due to their broad applicability across various material platforms and their straightforward fabrication process. This has enabled the demonstration of numerous applications on the SiN platform, including implantable nanophotonic neural probes for brain activity mapping \cite{brainmappingXue2024}, optical phased array beam scanners for LiDAR systems \cite{LidarBhandari2022}, and neuromorphic computing \cite{TOPSneuralnet}.

However, the TO effect in \( \text{Si}_3\text{N}_4 \) poses certain challenges, particularly due to its TO coefficient being an order of magnitude lower than that of silicon \cite{ArbabiTOSiN}, necessitating high-power inputs on the scale of several tens of milliwatts to achieve significant tunability \cite{SiNmW}, which in turn adversely impacts the response speed of the devices. Despite reported attempts to reduce power consumption \cite{submilliwatt}, the response time of these devices, often neglected, remains in the order of tens of microseconds \cite{brainSi3n4TO,SRN_TO}. 

In this context, achieving efficient temperature control and ensuring rapid response necessitates proximity between the metallic microheater and the photonic device to be tuned. However, the proximity introduces significant optical propagation losses in the device. A potential solution lies in leveraging parity-time (PT) symmetry breaking, a concept already explored in literature for silicon photonic waveguides \cite{Dave:19, PTcoupledmode, MartinezdeAguirreJokisch:24}, where strategically positioned metallic cladding directly atop the waveguide allows low-loss light propagation. Yet, to our knowledge, this approach remains unexplored for the silicon nitride-on-insulator platform.

This study presents a proof of concept for developing fast-responding TO modulators on \(\text{Si}_3\text{N}_4\) utilizing PT-symmetry breaking. In the first step, a low-loss, partially metal-clad waveguide operating at both 775 nm and 1550 nm wavelengths is optimized via FEM simulations. Subsequently, thermo-electrical simulations are employed to evaluate the temperature profile of the waveguide and the resulting phase shift corresponding to the applied electrical power on the microheater. Based on these predictions, the waveguide and the heater were incorporated into a Mach-Zehnder interferometer (MZI) to characterize the temperature-induced phase shift. While simulations were performed at 775 nm and 1550 nm wavelengths to demonstrate broadband operability, the fabricated device was characterized at the latter. Further, to highlight the advantages of our innovative heater design, we compare its performance with a conventional TO phase shifter design in which the microheater is placed atop an oxide cladding. This comparison underscores the improvements in power efficiency and response time achieved with our new design compared to the typical approach.

\section{Methods}
\subsection{PT-symmetry breaking in photonic waveguides}

\begin{figure}[ht!]
\centering\includegraphics[width=13cm]{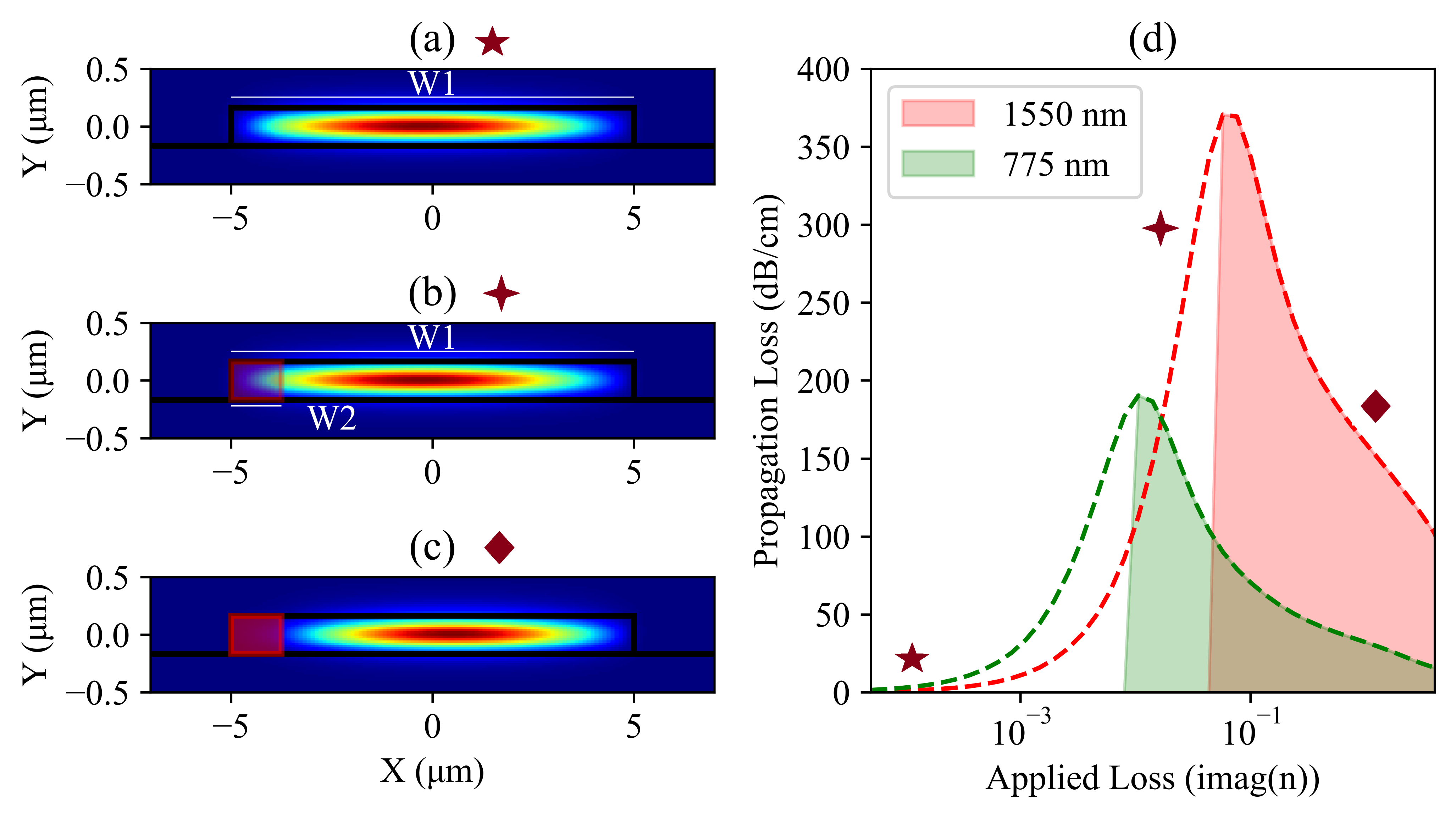}
\caption{Simulation of the PT-symmetry breaking in the photonic waveguide. (a) Fundamental mode profile (1550 nm) in a 6 $\mu m$ wide $\text{Si}_3\text{N}_4$ waveguide. (b) The same waveguide with a lossy section is indicated in red. The amount of loss applied is lower than the critical loss required for PT-symmetry breaking. (c) The PT-symmetry broken regime, in which the mode propagates with low loss confined away from the lossy section of the waveguide. (d) Propagation loss as a function of applied loss for different lossy section widths (W2) for 1550 nm and 775 nm wavelengths. The applied loss is manipulated by locally varying the imaginary part of the refractive index of the material. The filled regions of the curve depict the PT-symmetry broken regime.}
\label{fig:ptsym}
\end{figure}

As mentioned before, the primary limitation in designing reconfigurable thermo-optical devices is that achieving high efficiency requires positioning the heater in close proximity to the waveguide. However, this strategy inevitably leads to increased propagation losses. In our approach, we overcome this limitation by adopting the concept of parity-time (PT) symmetry breaking. When PT-symmetry is broken, the system can support modes where light can propagate with minimal loss despite the presence of absorptive materials, such as metals used for heating. 

To demonstrate the potential of PT-symmetry breaking in photonic waveguides, we simulate a \(\text{Si}_3\text{N}_4\) optical waveguide with a width of 6 $\mu$m, where a partial section (\(W2\)) of the same waveguide is made lossy by arbitrarily manipulating the imaginary part of the material's refractive index. This approach provides a clear and controlled way to initially investigate the effects of PT-symmetry breaking without the added complexity of integrating a metal layer.

Eigenmode simulations at 1550 nm wavelength estimated the propagation losses and electric field profiles of resonant modes. As illustrated in Figure \ref{fig:ptsym} (d), the propagation loss of the fundamental TE0 mode increases rapidly with increasing applied loss. Upon reaching a critical threshold, a new spatial mode emerges in the lossless region of the waveguide, entering the PT-symmetry broken regime, as shown in Figure \ref{fig:ptsym} (c). This new mode resembles the mode profile of the original TE0 mode (Figure \ref{fig:ptsym} (a)) but is confined to the lossless region. Further increases in applied loss result in reduced propagation loss of the mode due to stronger confinement away from the lossy region. To demonstrate the broadband applicability of PT-symmetry breaking, we showcase this phenomenon at 775 nm, taking full advantage of \(\text{Si}_3\text{N}_4\)'s transparency at this wavelength. As shown in Figure \ref{fig:ptsym}(d), the threshold loss for PT-symmetry breaking is lower at 775 nm, likely due to higher mode confinement in the waveguide.

\begin{figure}[ht!]
\centering\includegraphics[width=13cm]{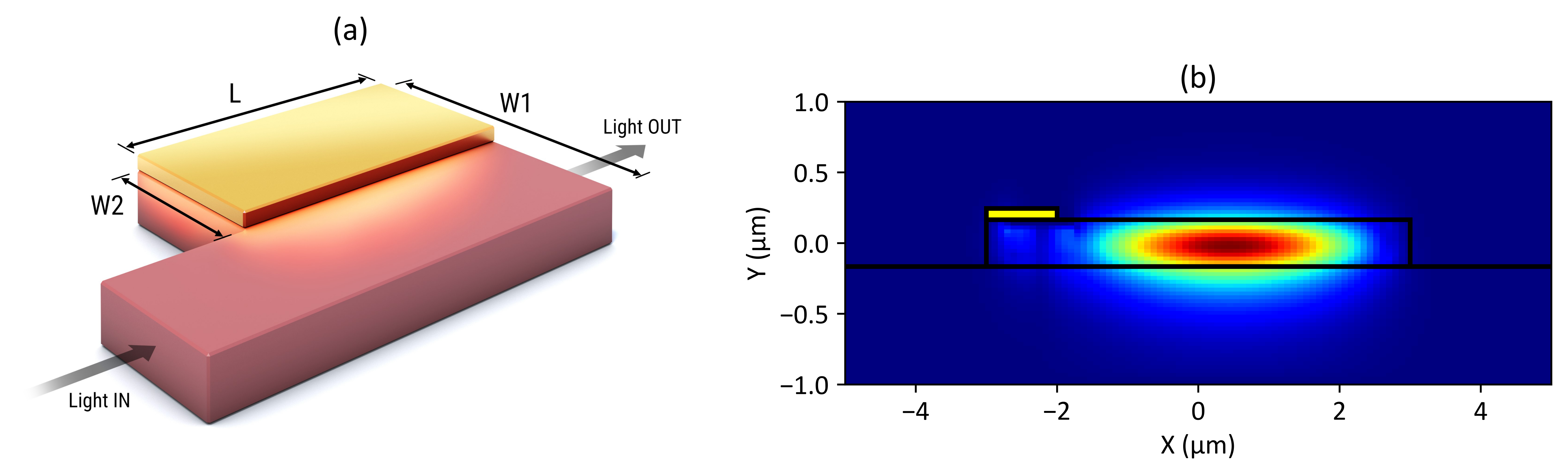}

\caption{ (a) A graphical render of the PT-symmetry breaking arrangement. A waveguide of width $W1$ is partially covered with $(W2)$ wide and 80 nm thin film of gold. The input waveguide with width $(W1-W2)$ excites the PT-symmetry broken mode which propagates over a distance L with low loss confined to the waveguide section without metal cladding. The proximity of the metal to the waveguide allows rapid temperature manipulation with Joule heating. (b) The propagating electric field profile at 1550 nm inside the PT-symmetry broken waveguide of 6 $\mu$m waveguide width with 1 $\mu$m wide gold microheater on top of it.}

\label{fig:render}
\end{figure}

The same phenomenon can be induced in the waveguide using a partial metal cladding, as shown in Figure \ref{fig:render} (a), instead of incorporating a separate lossy medium. This cladding will also double as a heater that can rapidly manipulate the temperature profile to take advantage of the TO effect. The simulated electric field profile of the fundamental TE0 mode for an 80 nm thick gold cladding is shown in Figure \ref{fig:render} (b). The propagation loss is estimated to be as low as 0.6 dB for a 100 $\mu$m long section of the waveguide, not accounting for potential scattering losses from fabrication imperfections.

\subsection{Thermo-optic phase shifter layout}

 To engineer the layout of our structures, we begin our simulations by modeling the temperature change ($\Delta T$) in the heated section of the waveguide and the corresponding effect on the refractive index ($\frac{dn}{dT}$) of the photonic material. We can then determine the induced temperature-dependent phase shift $\Delta \phi$ \cite{SiNpowerandrisereview}:
\begin{equation}
    \Delta \phi = \frac{2\pi L}{\lambda_0} \frac{dn}{dT} \Delta T
    \label{eqn:phi}
\end{equation}
Where $L$ is the length of the heated section, $\lambda_0$ is the wavelength of the light, and $\frac{dn}{dT}$ is the TO coefficient of the material at $T = 300 K$. For \( \text{Si}_3\text{N}_4 \), $\frac{dn}{dT}$ typically falls within the range of $2.5 \cdot 10^{-5}K^{-1}$ \cite{brainSi3n4TO}.

In our approach, we use a gold thin film deposited atop the waveguide as resistive metallic microheater, which serves to induce the required temperature change. For a microheater with resistance $R$ and current $I$ flowing through it, the power dissipated $P$ in the microheater can be determined using the Joule heating law: $P = RI^2$. The dissipated power generates heat, leading to a temperature change $\Delta T$, which can be calculated as $\Delta T = \frac{P}{G}$, where $G$ represents the thermal conductance between the heated waveguide and the surrounding sink. While these analytical equations offer an approximate estimate of $\Delta T$, they do not account for thermal boundary conductances in between materials. Therefore, the numerical finite element method (FEM) was employed to estimate the temperature profile of the photonic waveguide with better accuracy.

Once the temperature profile of the waveguide is obtained, the spatial variation of the refractive index is determined by calculating the perturbations at discrete 10 nm grid cells using the TO coefficient. Finite element eigenmode analysis is then performed yielding to the effective refractive index ($n_{eff} + ik_{eff}$) and the loss for the propagating mode. These simulation procedures are repeated, at the wavelength and device geometry of interest, for different applied microheater powers to calculate the resulting phase shift corresponding to the physical length ($L$) of the heating section (Figure \ref{fig:render} (a)). This procedure was repeated to estimate the performance metrics of a conventional heater-waveguide geometry, where the heater is placed above an 800 nm oxide cladding for a direct comparison when necessary. 

\subsection{Fabrication and characterization setup}

\begin{figure}[ht!]
\centering\includegraphics[width=13cm]{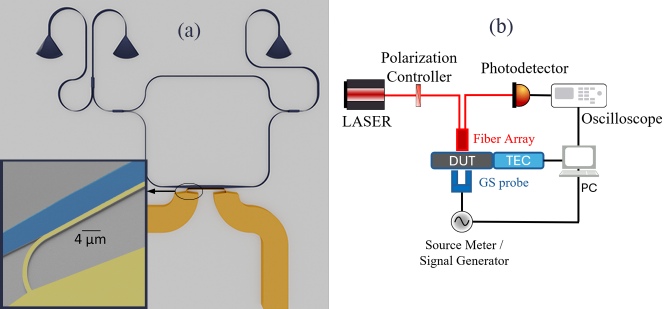}
\caption{(a) A graphical render of the phase shifter circuit. Light is coupled in and out of the device using apodized grating couplers. The input is split and later combined with 1x2 MMIs. The inset image shows a high-resolution micrograph acquired via scanning electron microscope of the metal (false-colored in yellow)  and waveguide (false-colored blue) interaction region of the fabricated device. (b) Schematic of the characterization setup.}
\label{fig:FabSEM}
\end{figure}

To experimentally characterize the phase shift induced by the microheater as a function of the applied electric power, the PT-symmetry broken waveguide was integrated into an unbalanced MZI, as shown in Figure \ref{fig:FabSEM} (a). Light is coupled into the device through an 8-degree fiber array focused onto an on-chip negative angle apodized grating coupler. A reference output grating coupler is also incorporated for alignment. The remaining light is split into two arms and recombined using symmetric Multimode Interferometers (MMIs). The unbalanced arm of the MZI includes a waveguide of width $W1$ = 6 $\mu$m and a heater of width $W2$ = 1 $\mu$m. The heated section is 1 mm long. Additionally, a fixed optical path difference of 100 $\mu$m compared to the second arm ensures an adequate free spectral range (FSR) to determine the thermo-optic phase shift. Finally, the output light from the MZI is coupled back into the fiber array through another grating coupler to measure the transmission through the device.

The device has been fabricated using a two-step electron-beam lithography (EBL) (Raith EBPG 5150) process. The substrate material stack wafer is comprised of low-stress 330 nm thick $\text{Si}_3\text{N}_4$ atop 3300 nm buried oxide on bulk Si. In Step 1, EBL is performed to define the markers for subsequent alignment, the microheater section, and the contact pads. We used CSAR 6200.13 e-beam positive resist for this step which was later developed with AR 600-546 developer. A 5 nm Cr layer deposited via physical vapor deposition (PVD), serves as an adhesion layer for the 80 nm layer of gold deposited on the chip. A Lift-off process is then executed using AR 600-71 CSAR remover.  Step 2 involves patterning the photonic components on the chip. Here, AR-N 7520.12 negative e-beam resist is used. After the lithography step, the waveguides are etched into the $\text{Si}_3\text{N}_4$ layer using a $CHF_3/O_2$ plasma. Finally, the resist is stripped with $O_2$ plasma. 

The characterization setup is shown in Figure \ref{fig:FabSEM}(b). The chip is mounted on a thermo-electric cooler (TEC) that stabilizes the substrate at a constant temperature of 25$^{\circ} C$ throughout the experiment. This setup ensures that potential fluctuations in optical properties due to ambient temperature variations are minimized. Additionally, the room is equipped with a closed controlled airflow system, which further mitigates the likelihood of significant changes in ambient conditions. The TEC is mounted on a 3-axis translation and rotational stage to allow precise alignment of the grating couplers with the fiber array used for optical access. The optical input signal is generated by a continuous wavelength (CW) tunable laser (Santec TSL 550) and the output signal is detected using a InGaAs photoreceiver (Newport 2053-FC-M) which is then digitised for further signal processing. The electrical signal is applied to the contact pads through a microprobe (Form Factor Cascade Unity GS Probe) mounted on a separate 3-axis translation stage for independent positioning. The probe is connected to a calibrated voltage source (Keithley 2450) to activate the heater.

\section{Results and Discussion}
\subsection{Static Performance}

\begin{figure}[ht!]
\centering\includegraphics[width=13cm]{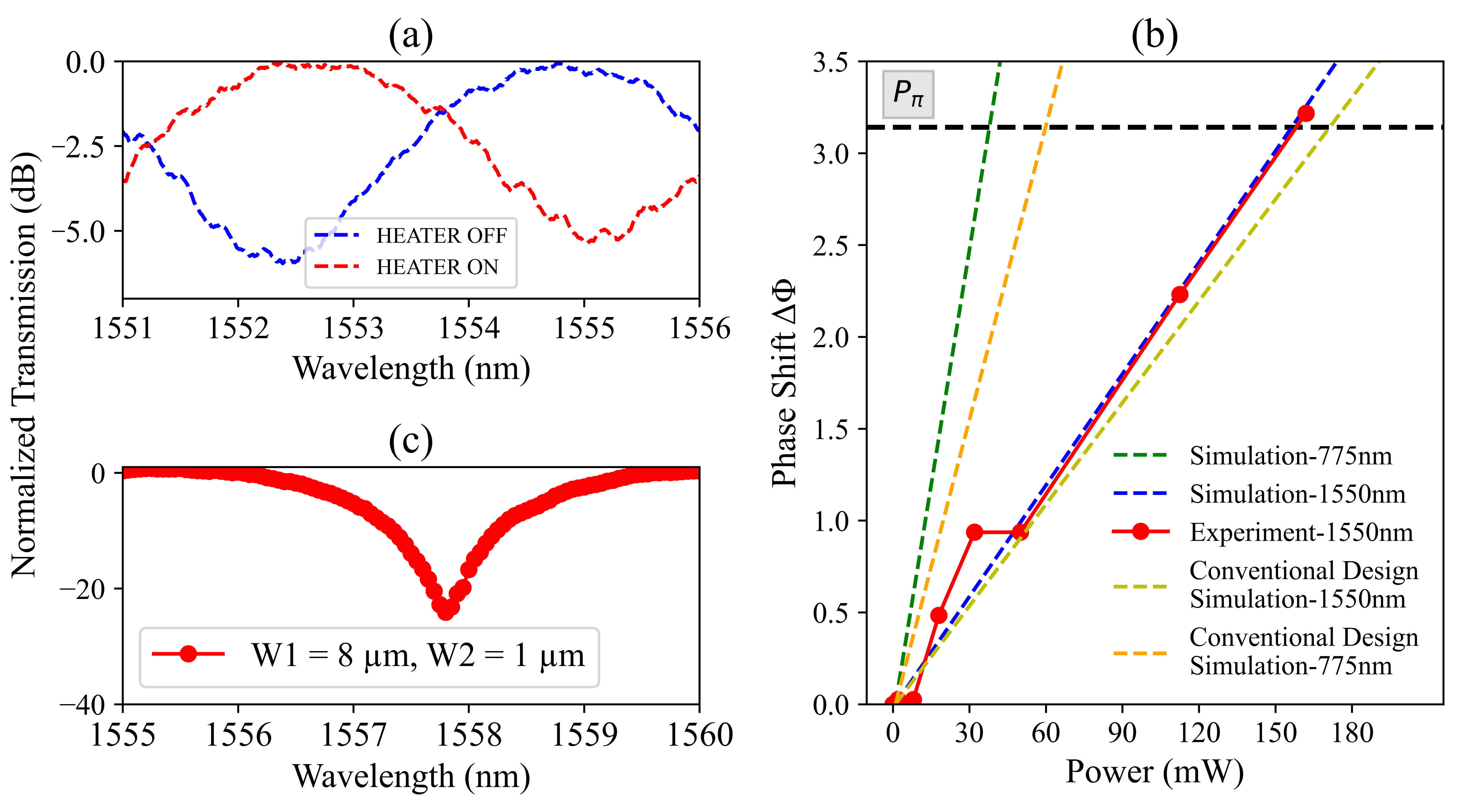}
\caption{Optical performance metrics of the TO phase shifter. (a) The optical transmission spectra before and after activating the microheater. (b) Shift in phase as a function of applied power on the microheater. The black line marks a $\pi$ shift for the device under consideration. (c) The measured optical transmission through a phase shifter with a waveguide of 8 $\mu$m width and microheater of 1 $\mu$m width. The propagation loss deduced from the ER of this geometry is 3 times lower compared to the previously presented 6 $\mu$m wide waveguide. }
\label{fig:Device23_plot}
\end{figure}

We first characterized our device in the steady state to investigate the power required to achieve a $\pi$ phase shift. With the phase shifter design with length $L = 1$ mm, an FSR of 5 nm was measured around 1550 nm and an extinction ratio (ER) of 7 dB, corresponding to a propagation loss of 8.3 dB/mm, which is higher than the simulated value of 5.7 dB/mm. This discrepancy is attributed to potential higher-order mode excitation in the wide waveguide which is eventually damped out due to our absorbing boundary. Later in this section, we propose a method to improve the propagation loss in our waveguide. By activating the microheater, we observed a red shift in the MZI fringes, as shown in Figure \ref{fig:Device23_plot} (a). A $\pi$ phase shift was achieved by applying 157 mW of power ($P_\pi$), which is in good agreement with the 156 mW predicted by our simulations (Figure \ref{fig:Device23_plot} (b)). Additionally, our simulations predict $P_\pi = 38$ mW when the same device is operated at a wavelength of 775 nm.

For a direct comparison, we also simulated a conventional TO phase shifter geometry. In this configuration, the waveguide was 1.1 $\mu$m wide with an 800 nm thick silicon dioxide cladding, and the heater was placed on top of the cladding. The length, width, and material of choice for the heater were identical to our new design. The $P_\pi$ for this conventional design was found to be considerably higher, at 171 mW and 60 mW for the 1550 nm and 775 nm wavelengths, respectively (Figure \ref{fig:Device23_plot} (c)). Although lower power consumption has been demonstrated in the literature, these results pertain to lower wavelengths \cite{brainSi3n4TO}, optimized heater materials \cite{SiNpowerandrisereview}, or folded multipass structures \cite{Ovvyan_2016}. Given these considerations, our design already exhibits improved power efficiency with potential for further optimization.

Furthermore, to reduce the propagation losses in the active area, we varied the width of the waveguide in the microheater region. We realized a set of test devices with different waveguide widths and microheater widths with $L = 100$ $\mu$m. For an 8 $\mu$m wide waveguide section with a 1 $\mu$m wide microheater, we measured a propagation loss of 3.9 dB/mm, significantly reducing the losses compared to a 6 $\mu$m wide waveguide. The reduction in losses significantly improved the measured ER from 7 dB to 25 dB, as shown in Figure \ref{fig:Device23_plot} (c). However, due to the shorter length of these devices and the relatively low temperature tolerance of the gold thin film, the heaters sustained damage before reaching the temperature required to induce a $\pi$ shift. Consequently, the same TO characterizations could not be performed on these devices. A more detailed discussion on the selection of the heater material will be provided in the next section.

\subsection{Response Time}
\begin{figure}[ht!]
\centering
\includegraphics[width=13cm]{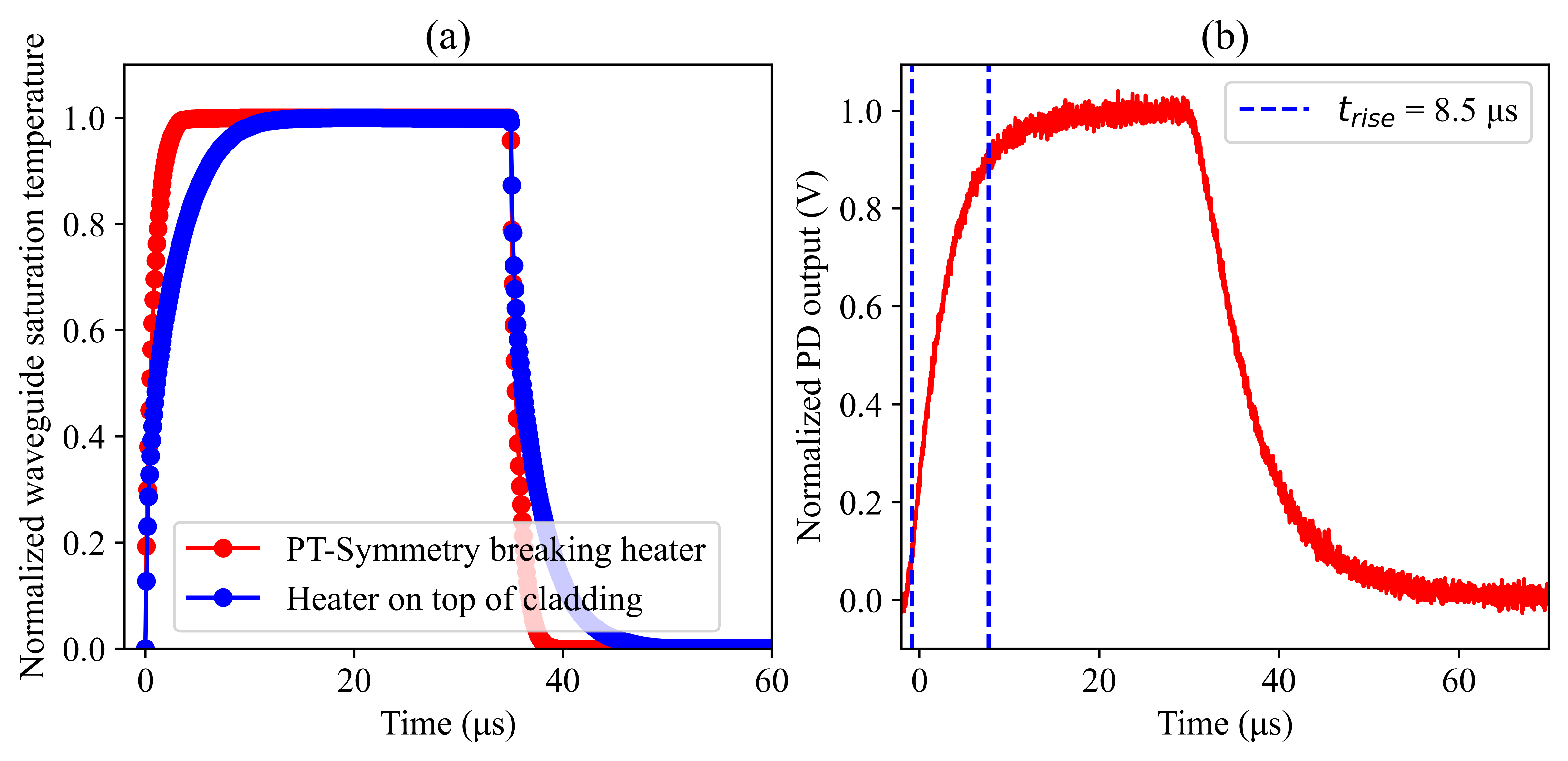}
\caption{Response time of the TO phase shifter. (a) Simulated temperature transient of a conventional microheater where the waveguide is 1.1 $\mu$m, the heater 1 $\mu$m wide on top of a cladding 1 $\mu$m thick (blue curve), and the same for the newly proposed PT-symmetry breaking microheater ($W1 = 6 \mu$m, $W2 = 1 \mu$m) (red curve). The rise and fall times are over 3 times shorter for the latter. (b) The measured thermo-optic timing response at 1550 nm for our PT-symmetry breaking TO phase shifter. The measured rise time of 8.5 $\mu$s, which to our knowledge, is among the fastest reported for the $\text{Si}_3\text{N}_4$ material platform.}
\label{fig:response_plot}
\end{figure}

Another important advantage of placing the microheater near the waveguide is the reduced time required for heat transfer, resulting in quicker TO response times. To support this hypothesis, heat transient simulations were performed for the newly proposed device geometry and compared with the conventional design mentioned previously. Figure \ref{fig:response_plot} (a) shows the temperature response in the waveguide obtained by applying a heating pulse to these geometries. Since the effective volume of the waveguides varies between the designs, a parameter called normalized waveguide saturation temperature is defined, representing the fraction of the average steady-state temperature obtainable in the $\text{Si}_3\text{N}_4$ waveguide for an arbitrary power applied to the microheater. Simulation results predict that the new design can achieve rise and fall times three times shorter than those of a conventional microheater geometry.

In our experiments with the device under investigation, we observed that a voltage pulse with an amplitude of 12 V was sufficient to induce a phase shift of \(\pi\) at 1550 nm, as shown in Figure \ref{fig:response_plot} (b). The measured rise time was 8.5 \(\mu\)s, defined as the time required for the signal to increase from 10\% to 90\% of its maximum amplitude. To the best of our knowledge, this rise time is among the fastest reported for TO phase shifters on the (\(\text{Si}_3\text{N}_4\)) material platform \cite{brainSi3n4TO, SiNpowerandrisereview, submilliwatt, SRN_TO}. These measurements confirm the enhanced response speed of our new design, in agreement with our initial simulations.

\subsection{Microheater Material}

The choice of material for the microheater is crucial for optimizing both the response time and power efficiency of thermo-optic phase shifters. Thermal conductivity is the most significant factor; higher thermal conductivity results in a shorter response time, while lower conductivity concentrates heat, leading to higher temperatures and a larger phase shift for the same power \cite{microheater_review}. Our choice of gold microheaters, which have excellent thermal conductivity (approximately 150 W/mK) \cite{gold_cond}, simplified the fabrication process and enabled fast response times but left room for improvement in the power efficiency of the device. Additionally, the structural rigidity of the gold microheater is compromised at high temperatures, as seen in the previous section, necessitating longer \cite{Ovvyan_2016} and consequently more lossy heating sections. Hence adopting heater materials with higher temperature tolerance, such as Titanium \cite{AuTi_paper}, is crucial for developing compact devices. Moreover, the gold microheaters likely suffer from increased resistance in response to rising temperatures, diminishing the dissipated power. This effect can be mitigated by using materials with a low temperature coefficient of electrical resistance, ensuring more stable performance. Therefore, switching to Titanium (Ti) or Titanium Nitride (TiN) microheaters, while maintaining gold contact pads for their favorable properties, could improve the power efficiency \cite{TiN} of our device. These alternatives ensure higher temperatures for any specific applied voltage and withstand temperatures up to 300°C, providing a balance between efficient heat transfer and improved performance metrics, as supported by existing literature \cite{microheater_review}.

\subsection{Thermal Cross-Talk}

\begin{figure}[ht!]
\centering
\includegraphics[width=13cm]{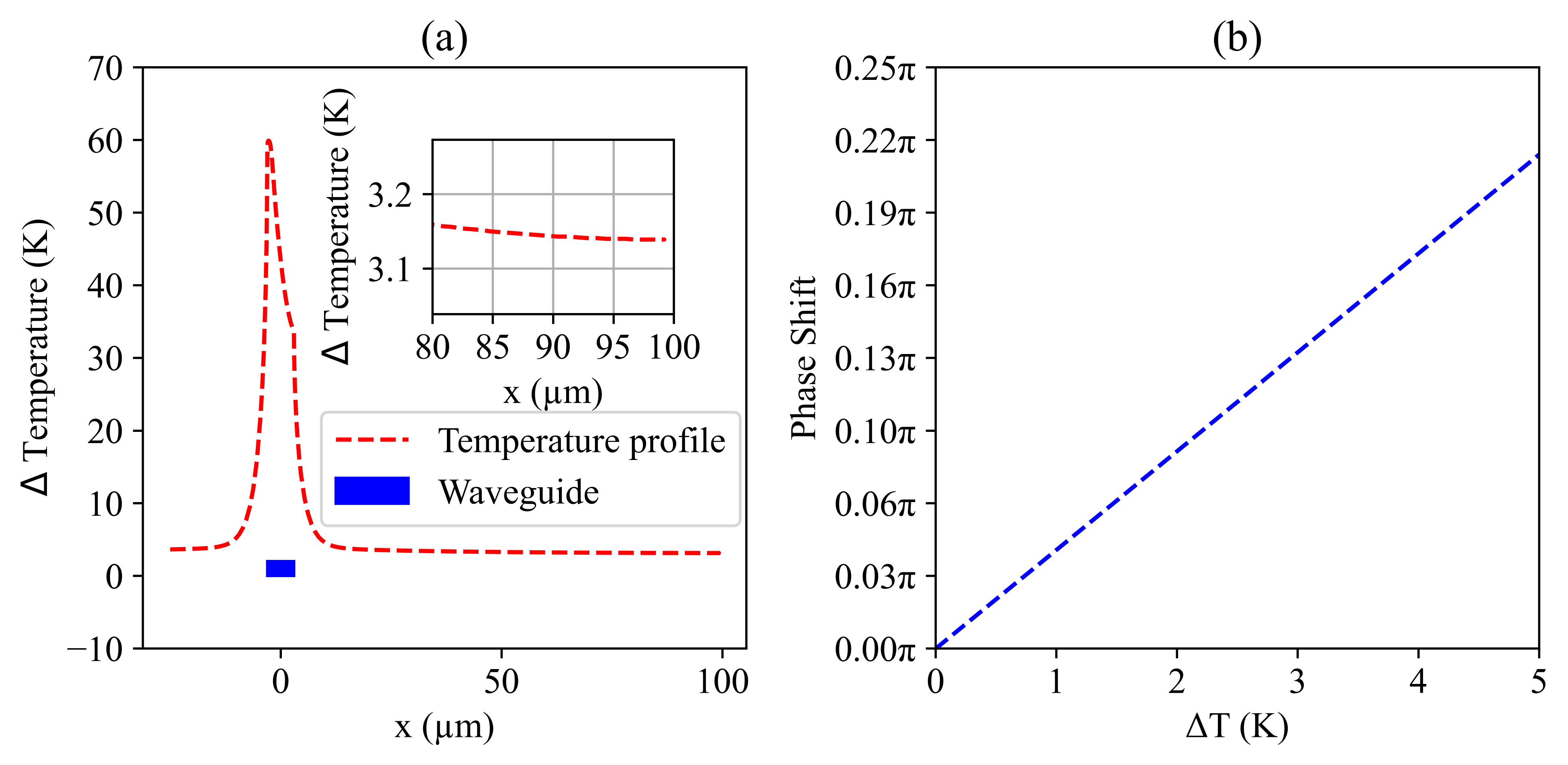}
\caption{(a) Simulated temperature difference over the bottom oxide surface around the heated waveguide. The zoomed inset shows the increase in temperature approximately 100 $\mu$m away. (b) The temperature-dependent phase shift generated in a parallel waveguide with the same length as the heated waveguide placed 100 $\mu$m away from the latter.}
\label{fig:cross_talk}
\end{figure}

We investigated the potential parasitic effects due to thermal cross-talk between the arms of the MZI. Figure \ref{fig:cross_talk} (a) shows the temperature profile for a waveguide centered at \( x = 0 \) at \( P_\pi \) = 150 mW according to our FEM simulations. The applied power in steady-state condition increases the bottom oxide surface temperature by 3 K even 100 \(\mu\)m away from the waveguide (Figure inset). This temperature rise induces a counteracting phase shift in the second arm of the same MZI, assuming it to be parallel and of the same length as the heated waveguide, as depicted in Figure \ref{fig:cross_talk} (b). The parasitic phase shift not only limits the ER of the device being tuned but also perturbs the working wavelength of neighboring devices.

Thermal cross-talk is a well-known effect in TO devices. Although the lower TO coefficient of \(\text{Si}_3\text{N}_4\) results in reduced instabilities compared to silicon, managing this effect is crucial for maximizing performance and enabling dense integration. To mitigate this one approach involves implementing heaters on both arms to enable both red and blue shifts of the MZI fringes \cite{Ovvyan_2016}, which corrects for any offsets caused by adjacent devices and allows for more independent operation. Additionally, various algorithms for thermal cross-talk correction in phase shifter arrays have been proposed, including those based on eigenmode decomposition \cite{crossalg1} and matrix inversion \cite{cross_alg2}. However, both of these methods require optical feedback loops and additional control electronics. A third and more robust strategy would be to implement deep-etched trenches between the arms \cite{crosstalktrench}. This method has been shown to manage the parasitic effects of thermal cross-talk effectively while maintaining CMOS compatibility of the fabrication process. Thermal insulation using trenches will also potentially improve the power efficiency of the phase shifter \cite{submilliwatt} by concentrating heat locally at the waveguide, resulting in higher temperatures for a given applied power.

\section{Conclusion}

This work demonstrates a proof of concept for fast-responding TO phase shifters on the \(\text{Si}_3\text{N}_4\) on insulator platform. Leveraging PT-symmetry breaking, a metallic microheater placed in direct contact with the waveguide allows rapid temperature control, achieving a notable thermo-optic response time of 8.5 \(\mu\)s for a complete switch of the output transmission. Despite the need to optimize the microheater material for improved power efficiency, the current device exhibits promising performance metrics and minimal propagation losses, with a 0.39 dB loss over a 100-\(\mu\)m-long heater-waveguide section. The two-step fabrication process further simplifies large-scale integration by obviating the need for an oxide cladding. 

Future improvements will focus on optimizing the heater material to enhance power efficiency and potentially reduce propagation losses, depending on the material’s absorption coefficient and the working wavelength. Given the heater's proximity to the waveguide, alternative materials with lower thermal conductivity are unlikely to impact response time significantly. Additionally, widening the waveguides may further decrease optical losses. Implementing cross-talk mitigation strategies suggested from the literature would also help realize dense arrangements of the phase shifter where necessary. Collectively, these advancements are expected to result in thermo-optic modulators with low losses, competitive response times, and straightforward fabrication, positioning them as a robust solution for advanced photonic applications.

Moreover, the metal-clad design is not confined to MZIs. Various photonic circuits optimized for different applications, as documented in the literature \cite{HybridINPMZI, racetrackSiN, SiN_MMI_MMZI, SiN_MRR}, can incorporate the same heater-waveguide geometry to enhance their response time while ensuring efficient heating. Although the demonstrated working wavelength in this study is 1550 nm and 775 nm, it is noteworthy that the TO effect on the silicon nitride on insulator platform can be exploited across its wide transparency window, catering to diverse applications. The experimental results presented here, along with the predictions supported by simulations and existing literature, underscore the promising potential of PT-symmetry breaking approach towards advancements in TO phase shifter technology on this material platform.

\printbibliography 

@Article{submilliwatt,
AUTHOR = {Wu, Zhaoyang and Lin, Shuqing and Yu, Siyuan and Zhang, Yanfeng},
TITLE = {Submilliwatt Silicon Nitride Thermo-Optic Modulator Operating at 532 nm},
JOURNAL = {Photonics},
VOLUME = {11},
YEAR = {2024},
NUMBER = {3},
ARTICLE-NUMBER = {213},
URL = {https://www.mdpi.com/2304-6732/11/3/213},
ISSN = {2304-6732},
DOI = {10.3390/photonics11030213}
}

@article{ultralowloss,
author = {Ji, Xingchen and Okawachi, Yoshitomo and Gil-Molina, Andres and Corato-Zanarella, Mateus and Roberts, Samantha and Gaeta, Alexander L. and Lipson, Michal},
title = {Ultra-Low-Loss Silicon Nitride Photonics Based on Deposited Films Compatible with Foundries},
journal = {Laser \& Photonics Reviews},
volume = {17},
number = {3},
pages = {2200544},
doi = {https://doi.org/10.1002/lpor.202200544},
url = {https://onlinelibrary.wiley.com/doi/abs/10.1002/lpor.202200544},
eprint = {https://onlinelibrary.wiley.com/doi/pdf/10.1002/lpor.202200544},
year = {2023}
}

@article{SiN_vis,
  title={Silicon Nitride Photonic Integration Platforms for Visible, Near-Infrared and Mid-Infrared Applications},
  author={Pascual Mu{\~n}oz and Gloria Mic{\'o} and Luis Alberto Bru and Daniel Pastor and Daniel P{\'e}rez and Jos{\'e} David Dom{\'e}nech and Juan Fern{\'a}ndez Salas and Roc{\'i}o Ba{\~n}os and Bernardo Gargallo and Rub{\'e}n Alemany and Ana M. S{\'a}nchez and Josep M. Cirera and Roser Mas and Carlos Dom{\'i}nguez},
  journal={Sensors (Basel, Switzerland)},
  year={2017},
  volume={17},
  url={https://api.semanticscholar.org/CorpusID:3516512}
}

@unknown{PTcoupledmode,
author = {Wei, Yanxian and Cheng, Junwei and Wang, Yilun and Zhou, Hailong and Dong, Jianji and Huang, Dongmei and Li, Feng and Li, Ming and Wai, P.K.A. and Zhang, Xinliang},
year = {2021},
month = {06},
pages = {},
title = {Fast-response silicon photonic microheater induced by parity-time symmetry breaking}
}

@Inbook{doping,
author="Liao, Ling
and Liu, Ansheng
and Nguyen, Hat
and Basak, Juthika
and Paniccia, Mario
and Chetrit, Yoel
and Rubin, Doron",
editor="Lockwood, David J.
and Pavesi, Lorenzo",
title="High-Speed Photonic Integrated Chip on a Silicon Platform",
bookTitle="Silicon Photonics II: Components and Integration",
year="2011",
publisher="Springer Berlin Heidelberg",
address="Berlin, Heidelberg",
pages="169--186",
isbn="978-3-642-10506-7",
doi="10.1007/978-3-642-10506-7_7",
url="https://doi.org/10.1007/978-3-642-10506-7_7"
}

@inproceedings{Dave:19,
author = {Utsav D. Dave and Michal Lipson},
booktitle = {Conference on Lasers and Electro-Optics},
journal = {Conference on Lasers and Electro-Optics},
pages = {FW4D.4},
publisher = {Optica Publishing Group},
title = {Low Loss Propagation in a Metal-clad Waveguide via PT-Symmetry Breaking},
year = {2019},
url = {https://opg.optica.org/abstract.cfm?URI=CLEO_QELS-2019-FW4D.4},
doi = {10.1364/CLEO_QELS.2019.FW4D.4},
}

@article{Ovvyan_2016,
doi = {10.1088/2040-8978/18/6/064011},
url = {https://dx.doi.org/10.1088/2040-8978/18/6/064011},
year = {2016},
month = {may},
publisher = {IOP Publishing},
volume = {18},
number = {6},
pages = {064011},
author = {A P Ovvyan and N Gruhler and S Ferrari and W H P Pernice},
title = {Cascaded Mach–Zehnder interferometer tunable filters},
journal = {Journal of Optics}
}

@article{microheater_review,
author={Jeroish, Z. E.
and Bhuvaneshwari, K. S.
and Samsuri, Fahmi
and Narayanamurthy, Vigneswaran},
title={Microheater: material, design, fabrication, temperature control, and applications---a role in COVID-19},
journal={Biomedical Microdevices},
year={2021},
month={Dec},
day={03},
volume={24},
number={1},
pages={3},
issn={1572-8781},
doi={10.1007/s10544-021-00595-8},
url={https://doi.org/10.1007/s10544-021-00595-8}
}

@article{SiNpowerandrisereview,
title = {Optimization and comprehensive comparison of thermo-optic phase shifter with folded waveguide on SiN and SOI platforms},
journal = {Optics Communications},
volume = {555},
pages = {130242},
year = {2024},
issn = {0030-4018},
doi = {https://doi.org/10.1016/j.optcom.2023.130242},
url = {https://www.sciencedirect.com/science/article/pii/S0030401823009902},
author = {Jin Wang and Wei Cheng and Wanghua Zhu and Mengjia Lu and Yifei Chen and Shangqing Shi and Chen Guo and Guohua Hu and Yiping Cui and Binfeng Yun}
}

@article{SiNmW,
author = {Min Chul Shin and Aseema Mohanty and Kyle Watson and Gaurang R. Bhatt and Christopher T. Phare and Steven A. Miller and Moshe Zadka and Brian S. Lee and Xingchen Ji and Ipshita Datta and Michal Lipson},
journal = {Opt. Lett.},
number = {7},
pages = {1934--1937},
publisher = {Optica Publishing Group},
title = {Chip-scale blue light phased array},
volume = {45},
month = {Apr},
year = {2020},
url = {https://opg.optica.org/ol/abstract.cfm?URI=ol-45-7-1934},
doi = {10.1364/OL.385201},
}

@article{SiN_MRR,
  title={A High-Performance Microwave Photonic Phase Shifter Based on Cascaded Silicon Nitride Microrings},
  author={Dongdong Lin and Xuemeng Xu and Pengfei Zheng and Guohua Hu and Binfeng Yun and Yiping Cui},
  journal={IEEE Photonics Technology Letters},
  year={2020},
  volume={32},
  pages={1265-1268},
  url={https://api.semanticscholar.org/CorpusID:221592298}
}

@article{brainSi3n4TO,
author = {Mohanty, Aseema and Li, Qian and Tadayon, Mohammad and Roberts, Samantha and Bhatt, Gaurang and Shim, Euijae and Ji, Xingchen and Cardenas, Jaime and Miller, Steven and Kepecs, Adam and Lipson, Michal},
year = {2020},
month = {02},
pages = {},
title = {Reconfigurable nanophotonic silicon probes for sub-millisecond deep-brain optical stimulation},
volume = {4},
journal = {Nature Biomedical Engineering},
doi = {10.1038/s41551-020-0516-y}
}

@article{SRN_TO,
author = {Hani Nejadriahi and Alex Friedman and Rajat Sharma and Steve Pappert and Yeshaiahu Fainman and Paul Yu},
journal = {Opt. Express},
number = {17},
pages = {24951--24960},
publisher = {Optica Publishing Group},
title = {Thermo-optic properties of silicon-rich silicon nitride for on-chip applications},
volume = {28},
month = {Aug},
year = {2020},
url = {https://opg.optica.org/oe/abstract.cfm?URI=oe-28-17-24951},
doi = {10.1364/OE.396969},
}

@article{Xiang:22,
author = {Chao Xiang and Warren Jin and John E. Bowers},
journal = {Photon. Res.},
number = {6},
pages = {A82--A96},
publisher = {Optica Publishing Group},
title = {Silicon nitride passive and active photonic integrated circuits: trends and prospects},
volume = {10},
month = {Jun},
year = {2022},
url = {https://opg.optica.org/prj/abstract.cfm?URI=prj-10-6-A82},
doi = {10.1364/PRJ.452936},
}

@article{ArbabiTOSiN,
author = {Arbabi, Amir and Goddard, Lynford},
year = {2013},
month = {09},
pages = {},
title = {Measurements of the refractive indices and thermo-optic coefficients of Si3N4 and SiO x using microring resonances},
volume = {38},
journal = {Optics Letters},
doi = {10.1364/OL.38.003878}
}

@article{TiN,
    author = {Singh, Surinder and Jejusaria, Alok and Singh, Jaspreet and Vashishath, Munish and Kumar, Dinesh},
    title = "{Comparative study of titanium, platinum, and titanium nitride thin films for micro-elecrto mechanical systems (MEMS) based micro-heaters}",
    journal = {AIP Advances},
    volume = {12},
    number = {9},
    pages = {095202},
    year = {2022},
    month = {09},
    issn = {2158-3226},
    doi = {10.1063/6.0001892},
    url = {https://doi.org/10.1063/6.0001892},
    eprint = {https://pubs.aip.org/aip/adv/article-pdf/doi/10.1063/6.0001892/16474183/095202\_1\_online.pdf},
}

@article{AuTi_paper,
title = {Fabrication, modeling and testing of a thin film Au/Ti microheater},
journal = {International Journal of Thermal Sciences},
volume = {46},
number = {6},
pages = {580-588},
year = {2007},
issn = {1290-0729},
doi = {https://doi.org/10.1016/j.ijthermalsci.2006.08.002},
url = {https://www.sciencedirect.com/science/article/pii/S1290072906001505},
author = {K.L. Zhang and S.K. Chou and S.S. Ang}
}

@article{gold_cond,
author = {Gilani, T.H. and Rabchuk, Dian},
title = {Electrical resistivity of gold thin film as a function of film thickness},
journal = {Canadian Journal of Physics},
volume = {96},
number = {3},
pages = {272-274},
year = {2018},
doi = {10.1139/cjp-2017-0484},

URL = { 
    
        https://doi.org/10.1139/cjp-2017-0484
    
    

},
eprint = { 
    
        https://doi.org/10.1139/cjp-2017-0484
    
    

}
}

@INPROCEEDINGS{SiN_MMI_MMZI,
  author={Rao, Ashutosh and Moille, Gregory and Lu, Xiyuan and Westly, Daron and Geiselmann, Michael and Zervas, Michael and Srinivasan, Kartik},
  booktitle={2021 Conference on Lasers and Electro-Optics (CLEO)}, 
  title={Up to 50 dB Extinction in Broadband Single-Stage Thermo-Optic Mach-Zehnder Interferometers for Programmable Low-Loss Silicon Nitride Photonic Circuits}, 
  year={2021},
  volume={},
  number={},
  pages={1-2},
  doi={}}

@article{Churaev2023,
author={Churaev, Mikhail
and Wang, Rui Ning
and Riedhauser, Annina
and Snigirev, Viacheslav
and Bl{\'e}sin, Terence
and M{\"o}hl, Charles
and Anderson, Miles H.
and Siddharth, Anat
and Popoff, Youri
and Drechsler, Ute
and Caimi, Daniele
and H{\"o}nl, Simon
and Riemensberger, Johann
and Liu, Junqiu
and Seidler, Paul
and Kippenberg, Tobias J.},
title={A heterogeneously integrated lithium niobate-on-silicon nitride photonic platform},
journal={Nature Communications},
year={2023},
month={Jun},
day={13},
volume={14},
number={1},
pages={3499},
issn={2041-1723},
doi={10.1038/s41467-023-39047-7},
url={https://doi.org/10.1038/s41467-023-39047-7}
}

@article{racetrackSiN,
author = {Abu Naim R. Ahmed and Shouyuan Shi and Andrew J. Mercante and Dennis W. Prather},
journal = {Opt. Express},
number = {21},
pages = {30741--30751},
publisher = {Optica Publishing Group},
title = {High-performance racetrack resonator in silicon nitride - thin film lithium niobate hybrid platform},
volume = {27},
month = {Oct},
year = {2019},
url = {https://opg.optica.org/oe/abstract.cfm?URI=oe-27-21-30741},
doi = {10.1364/OE.27.030741},
}

@article{HybridINPMZI,
author = {William M. J. Green and Reginald K. Lee and Guy A. DeRose and Axel Scherer and Amnon Yariv},
journal = {Opt. Express},
number = {5},
pages = {1651--1659},
publisher = {Optica Publishing Group},
title = {Hybrid InGaAsP-InP Mach-Zehnder Racetrack Resonator for Thermooptic Switching and Coupling Control},
volume = {13},
month = {Mar},
year = {2005},
url = {https://opg.optica.org/oe/abstract.cfm?URI=oe-13-5-1651},
doi = {10.1364/OPEX.13.001651},
}

@article{MartinezdeAguirreJokisch:24,
author = {Be\~{n}at Martinez de Aguirre Jokisch and Rasmus Elleb{\ae}k Christiansen and Ole Sigmund},
journal = {J. Opt. Soc. Am. B},
number = {2},
pages = {A18--A31},
publisher = {Optica Publishing Group},
title = {Topology optimization framework for designing efficient thermo-optical phase shifters},
volume = {41},
month = {Feb},
year = {2024},
url = {https://opg.optica.org/josab/abstract.cfm?URI=josab-41-2-A18},
doi = {10.1364/JOSAB.499979},
}

@article{brainmappingXue2024,
author={Xue, Tianyuan
and Stalmashonak, Andrei
and Chen, Fu-Der
and Ding, Peisheng
and Luo, Xianshu
and Chua, Hongyao
and Lo, Guo-Qiang
and Sacher, Wesley D.
and Poon, Joyce K. S.},
title={Implantable photonic neural probes with out-of-plane focusing grating emitters},
journal={Scientific Reports},
year={2024},
month={Jun},
day={15},
volume={14},
number={1},
pages={13812},
issn={2045-2322},
doi={10.1038/s41598-024-64037-0},
url={https://doi.org/10.1038/s41598-024-64037-0}
}

@article{LidarBhandari2022,
author={Bhandari, Bishal
and Wang, Chenxi
and Gwon, Ji-Yeong
and Heo, Jin-Moo
and Ko, Sung-Yong
and Oh, Min-Cheol
and Lee, Sang-Shin},
title={Dispersive silicon--nitride optical phased array incorporating arrayed waveguide delay lines for passive line beam scanning},
journal={Scientific Reports},
year={2022},
month={Nov},
day={05},
volume={12},
number={1},
pages={18759},
issn={2045-2322},
doi={10.1038/s41598-022-23456-7},
url={https://doi.org/10.1038/s41598-022-23456-7}
}

@inproceedings{TOPSneuralnet,
author = {H. H. Zhu and J. Zou and H. Zhang and H. Cai and A. Q. Liu},
booktitle = {Conference on Lasers and Electro-Optics},
journal = {Conference on Lasers and Electro-Optics},
pages = {SF1C.1},
publisher = {Optica Publishing Group},
title = {A Space-efficient Optical Computing Chip Based on Diffractive Neural Network},
year = {2022},
url = {https://opg.optica.org/abstract.cfm?URI=CLEO_SI-2022-SF1C.1},
doi = {10.1364/CLEO_SI.2022.SF1C.1},
}

@article{SiNLidar,
author = {Christopher V. Poulton and Matthew J. Byrd and Manan Raval and Zhan Su and Nanxi Li and Erman Timurdogan and Douglas Coolbaugh and Diedrik Vermeulen and Michael R. Watts},
journal = {Opt. Lett.},
number = {1},
pages = {21--24},
publisher = {Optica Publishing Group},
title = {Large-scale silicon nitride nanophotonic phased arrays at infrared and visible wavelengths},
volume = {42},
month = {Jan},
year = {2017},
url = {https://opg.optica.org/ol/abstract.cfm?URI=ol-42-1-21},
doi = {10.1364/OL.42.000021},
}

@article{SiNLifescience,
author = {Alireza Tabatabaei Mashayekh and Thomas Klos and Douwe Geuzebroek and Edwin Klein and Theo Veenstra and Martin B\"{u}scher and Florian Merget and Patrick Leisching and Jeremy Witzens},
journal = {Opt. Express},
number = {6},
pages = {8635--8653},
publisher = {Optica Publishing Group},
title = {Silicon nitride PIC-based multi-color laser engines for life science applications},
volume = {29},
month = {Mar},
year = {2021},
url = {https://opg.optica.org/oe/abstract.cfm?URI=oe-29-6-8635},
doi = {10.1364/OE.417245},
}

@INPROCEEDINGS{SiNTele,
  author={Maegami, Yuriko and Cong, Guangwei and Ohno, Morifumi and Narushima, Toshihiro and Yamamoto, Noritsugu and Kawashima, Hitoshi and Yamada, Koji},
  booktitle={2023 Opto-Electronics and Communications Conference (OECC)}, 
  title={Silicon-Nitride-based Passive Photonic Platform for Visible and Telecommunications Wavelength Regions}, 
  year={2023},
  volume={},
  number={},
  pages={1-3},
  doi={10.1109/OECC56963.2023.10209769}}

@article{crossalg1,
author = {Maziyar Milanizadeh and Douglas Aguiar and Andrea Melloni and Francesco Morichetti},
journal = {J. Lightwave Technol.},
number = {4},
pages = {1325--1332},
publisher = {Optica Publishing Group},
title = {Canceling Thermal Cross-Talk Effects in Photonic Integrated Circuits},
volume = {37},
month = {Feb},
year = {2019},
url = {https://opg.optica.org/jlt/abstract.cfm?URI=jlt-37-4-1325},
}

@article{cross_alg2,
  author={Milanizadeh, Maziyar and Ahmadi, Sara and Petrini, Matteo and Aguiar, Douglas and Mazzanti, Riccardo and Zanetto, Francesco and Guglielmi, Emanuele and Sampietro, Marco and Morichetti, Francesco and Melloni, Andrea},
  journal={IEEE Journal of Selected Topics in Quantum Electronics}, 
  title={Control and Calibration Recipes for Photonic Integrated Circuits}, 
  year={2020},
  volume={26},
  number={5},
  pages={1-10},
  doi={10.1109/JSTQE.2020.2975657}}

@article{crosstalktrench,
author = {Maxime Jacques and Alireza Samani and Eslam El-Fiky and David Patel and Zhenping Xing and David V. Plant},
journal = {Opt. Express},
number = {8},
pages = {10456--10471},
publisher = {Optica Publishing Group},
title = {Optimization of thermo-optic phase-shifter design and mitigation of thermal crosstalk on the SOI platform},
volume = {27},
month = {Apr},
year = {2019},
url = {https://opg.optica.org/oe/abstract.cfm?URI=oe-27-8-10456},
doi = {10.1364/OE.27.010456},
}

@article{sin1,
author = {Jared F. Bauters and Martijn J. R. Heck and Demis D. John and Jonathon S. Barton and Christiaan M. Bruinink and Arne Leinse and Ren\'{e} G. Heideman and Daniel J. Blumenthal and John E. Bowers},
journal = {Opt. Express},
number = {24},
pages = {24090--24101},
publisher = {Optica Publishing Group},
title = {Planar waveguides with less than 0.1 dB/m propagation loss fabricated with wafer bonding},
volume = {19},
month = {Nov},
year = {2011},
url = {https://opg.optica.org/oe/abstract.cfm?URI=oe-19-24-24090},
doi = {10.1364/OE.19.024090},
}

@article{sin2,
author = {Heck, Martijn J. R. and Bauters, Jared F. and Davenport, Michael L. and Spencer, Daryl T. and Bowers, John E.},
title = {Ultra-low loss waveguide platform and its integration with silicon photonics},
journal = {Laser \& Photonics Reviews},
volume = {8},
number = {5},
pages = {667-686},
doi = {https://doi.org/10.1002/lpor.201300183},
url = {https://onlinelibrary.wiley.com/doi/abs/10.1002/lpor.201300183},
eprint = {https://onlinelibrary.wiley.com/doi/pdf/10.1002/lpor.201300183},
year = {2014}
}

\end{document}